\documentclass{sig-alternate}
\usepackage{paralist}
\newcommand{\HASH}{\mathrm{H}}
\newcommand{\ENC}{\mathrm{E}}
\newcommand{\App}{{\ensuremath{\sf App}}}
\newcommand{\Fixt}{{\ensuremath{\sf Fix}}}
\newcommand{\CM}{{\ensuremath{\sf CM}}}
\newcommand{\AM}{{\ensuremath{\sf AM}}}
\newcommand{\Adv}{{{$\mathcal{ADV}$}}}
\usepackage{epsfig,endnotes}
\usepackage{amssymb, amsmath, latexsym, amsfonts, mathrsfs}
\usepackage{hyperref}
\usepackage{breakurl}
\usepackage{subfigure}
\usepackage{balance}
\usepackage{times}
\usepackage{microtype}
\usepackage{lscape}

\newcommand{\subparagraph}{}
\usepackage[compact]{titlesec}
\titlespacing*{\section}{0pt}{*2}{3pt}
\titlespacing{\subsection}{0pt}{*1.5}{2pt}
\titlespacing{\subsubsection}{0pt}{*1.5}{0pt}

\DeclareMathAlphabet{\mathcal}{OMS}{cmsy}{m}{n} 

\newcommand{\ignore}[1]{}
\begin{document}
\title{Securing Instrumented Environments over \\Content-Centric Networking:
the Case of Lighting Control}
\author{Jeff Burke \and Paolo Gasti \and Naveen Nathan \and Gene Tsudik}
\date{}

\makeatletter
\let\@copyrightspace\relax
\makeatother

\maketitle

\begin{abstract}
Instrumented environments, such as modern building automation systems (BAS), 
are becoming commonplace and are
increasingly interconnected with (and sometimes by) enterprise networks and the Internet. 
Regardless of the underlying communication platform, secure control of devices in such environments 
is a challenging task. The current trend is to move from proprietary communication media and protocols 
to IP over Ethernet. 
While the move to IP represents progress, new and different Internet architectures might
be better-suited for instrumented environments.

In this paper, we consider security of instrumented environments in the context of Content-Centric 
Networking (CCN). In particular, we focus on building automation over Named-Data Networking (NDN), a 
prominent instance of CCN. 
After identifying 
security requirements in a specific BAS sub-domain (lighting control), 
we construct a concrete NDN-based security architecture,
analyze its properties and report on preliminary implementation and experimental results.
We believe that this work represents a useful exercise in assessing the utility of NDN
in securing a communication paradigm well outside of its claimed {\em forte} of content distribution.  
At the same time, we provide a viable (secure and efficient) communication platform for a class
of instrumented environments exemplified by lighting control.
\end{abstract}

\section{Introduction}
The Internet has clearly proven to be a tremendous global success. Billions of people worldwide 
use it to perform a wide range of everyday tasks. It hosts a large number of information-intensive 
services, involves enormous amounts of content created and consumed over the Web, and
interconnects untold millions of wired, wireless, fixed and mobile computing devices.

Since Internet's inception, the amount of data exchanged over it has witnessed exponential growth.
Recently, this growth intensified due to increases in: (1) distribution of multimedia content, (2) popularity
of social networks and (3) amount of user-generated content. Unfortunately, the same usage model 
that fueled Internet's success is also exposing its limitations. Core ideas of today's Internet were 
developed in the 1970-s, when telephony -- i.e., a point-to-point conversation between two entities -- was 
the only successful example of effective global-scale communication technology. Moreover, original Internet
applications were few and modest in nature, e.g., store-and-forward email and remote computer access.

The world has changed dramatically since the 1970-s and the Internet now has to accommodate 
new services and applications as well as different usage models. To keep pace with changes and 
move the Internet into the future, several research efforts to design new Internet architectures have
been initiated in recent years.

Named-Data Networking (NDN) \cite{NDN} is an on-going research project that aims to 
develop a candidate next-generation Internet architecture. NDN exemplifies the so-called 
Content-Centric approach \cite{gritter2001architecture,Jacobson2009,koponen2007data} 
to networking. It explicitly names content instead of physical locations (i.e.,
hosts or network interfaces) and thus transforms content into a first-class entity. NDN also stipulates 
that each piece of named content must be digitally signed by its producer. This allows decoupling of 
trust in content from trust in the entity that might store and/or disseminate that content. These
NDN features facilitate automatic caching of content to optimize bandwidth use and enable 
effective simultaneous utilization of multiple network interfaces.

NDN's long-term goal is to replace TCP/IP. In order to succeed, NDN must prove that it can be 
used to efficiently implement all kinds of communication commonly performed over IP today and
envisaged for the near future. 
NDN has been shown as a viable architecture for content distribution \cite{Jacobson2009} 
as well as real-time \cite{VanSmBriPlStThBra09-Voice} and anonymous communication \cite{andana}. 
However, it remains unclear how NDN would fare in the context of other, less content-centric, 
communication paradigms, such as: cyber-physical systems (CPS), group communication 
(e.g., conferencing) and instrumented environments (e.g., building automation). 

\subsection{Building Automation and\\ Lighting Control} 
Building Automation Systems (BAS) provide a 
hardware and software platform for control, monitoring and management of:
heating, ventilation and air conditioning (HVAC), lighting, water, physical access control and other 
building components. BAS are subject to several important 
current trends with security implications:  
\begin{compactitem}
\item Increasing use of IP and Ethernet for industrial control, often over commodity wiring and network 
hardware.
\item Convergence of previously separate networks for automation and IT enabled 
by this new common infrastructure.
\item Increasing interest in cyber-physical systems (CPS) that leverage internetworking 
of physical and digital elements to enable and develop new types of applications.  
\end{compactitem}
In general, BAS offers an interesting and challenging application domain for 
NDN because content-centric networking  is generally \linebreak discussed in terms of
its improvements to content retrieval,
as opposed to control, actuation, or remote execution.
In the domain of BAS, we focus on lighting control as the initial test platform for the design and 
implementation of control communication over NDN. 
This choice is based on three reasons: (1) lighting represents a broad class of actuators 
while incurring limited physical safety concerns; (2) prevalence of IP-based control of lighting fixtures in 
new architectural and entertainment deployments; (3) access to comparisons with IP- and 
serially-controlled systems.   

In designing an NDN lighting control framework, our goals are:
\begin{compactenum}
\item Satisfy low latency requirements for communication between software (or hardware) 
controllers and lighting fixtures.
\item Use NDN content naming to address all components of the system, with names related to 
their identity or function rather than a combination of addressing that spans layers and systems 
(e.g., VLAN tag, IP address of gateway, port of protocol, address of fixture) as in current implementations.  
\item Given that 
NDN makes widespread use of content signatures, 
identify every entity in the system by a distinct public key.
\item Control access to fixtures via authorization policies, coupled with strong authentication.
(Current BAS and lighting systems typically rely on physical or VLAN-based segregation for security, 
making interoperability with IT systems challenging, configurations brittle to change, and requiring 
advanced networking expertise to set up and maintain.)
\item Use NDN naming itself to reflect access restrictions, rather than require a separate policy language.  
The main motivation is that a namespace is consistently accessible within any NDN-compliant device or process. 
This obviates the need for application-specific access control protocols. 
\item Develop security mechanisms suitable for low-power systems, initially targeting cell-phone class 
devices with a planned transition to microcontrollers typical of IP-connected lights today, such as the 
Phillips Color Kinetics ColorBlaze, that uses a 72-MHz ARM processor.  
\end {compactenum}

\subsection{Securing Lighting Control}  
Current lighting control installations, especially in non-critical facilities, tend to rely on 
network segregation and/or VLANs and VPNs to isolate their control traffic from 
general-purpose IP communication.
everal trends, such as increasing emphasis on energy management and the ``smart grid'' suggest
that this will not remain a viable approach in the future: Instrumented environments have 
increasing reliance on the 
Internet for patches and updates, remote access, data gathering, and application integration, as well as 
increasing opportunities for integration in homes and other environments without 
enterprise-level security. 
Consequently, we believe that, in the near future, it will become increasingly difficult to provide 
effective security by network separation alone: the 
most compelling applications depend on interfaces across system and network boundaries. 
VLANs, IP subnetting, and other network configurations spread addressing information 
across network layers in a way that is rarely meaningful to end-users or application developers.

Besides lower complexity and greater interoperability, running lighting control applications
over public networks brings certain advantages. First,  there is no need to design, deploy and manage a 
separate network infrastructure, since lighting control can benefit from high-speed, low-latency, fault-tolerant 
networks already deployed for general-purpose communication needs. Second, lighting control can be physically 
distributed, with devices spanning buildings and sites, and applications accessing them from a variety of 
locations.  Third, separate (often esoteric and proprietary) security measures common in 
today's lighting control would be unnecessary, due to availability of standardized security features and techniques.

\subsection{Focus} \label{sec:Focus}
This paper is focused on securing lighting control systems running over NDN. As mentioned above, 
allowing control messages to reach actual lighting fixtures (as opposed to dedicated controllers) 
imposes strict performance constraints, 
in addition to more general requirements of availability and fault tolerance. While general BAS might tolerate 
variable delays up to a few seconds for actuating or sensing slowly changing systems, latency requirements 
for lighting control are stricter and represent an overlap with industrial and process control. 
To provide a sense of ``real-time'' interaction, architectural lighting might require execution of commands 
within a few hundred milliseconds of pressing a switch, or 
updates close to 44Hz DMX refresh rate \cite{dmx} to achieve a smooth fade from one value to another.  
By designing and implementing a secure lighting framework suitable for such low latency systems coupled with
a meaningful namespace, we target a hybrid design space that corresponds to the so-called 
``thin waist'' for highly heterogenous BAS of the future. 

\smallskip\noindent
{\textbf{Organization:}}
We proceed with NDN overview in Section \ref{sec:ndn}. 
It is followed by the description of a base-line lighting control protocol in Section
\ref{sec:build-autom}. The same section introduces our framework. Implementation details and performance
evaluation results are discussed in Section \ref{sec:performance}. Section \ref{sec:security} contains the  security analysis.
Next, Section \ref{sec:relwork} summarizes related
work. The paper concludes with future work agenda and a summary
in Section \ref{sec:conclusions}.

\section{Overview of NDN}
\label{sec:ndn}
NDN \cite{NDN} is a communication architecture based on named content.
Rather than addressing content by its location, NDN refers to it by name. A content name is
composed of one or more variable-length components. 
Component boundaries are explicitly delimited by ``{{\small \tt
/}''. For example, the name of a CNN news content might be: {\small
\tt /ndn/cnn/news/2011aug20}. Large pieces of content can be split
into fragments with predictable names: fragment $137$ of a YouTube
video could be named: {\small \tt /ndn/youtube/video-749.avi/137}.

Since the main abstraction is content, there is no
explicit notion of ``hosts'' in NDN. (However, their existence is assumed.)
Communication adheres to the {\em pull} model: content is delivered
to consumers only upon explicit request. A consumer requests content
by sending an {\em interest} packet. If an entity (a router or a host)
can ``satisfy'' a given interest, it returns the corresponding {\em content object}. 
Interest and content are the only types of packets in NDN. A
content packet with name X in NDN is  never forwarded or routed
unless it is preceded by an interest for name X.\footnote{\scriptsize Strictly speaking,
content named X$'\neq$ X can be delivered in response to an interest
for X, but only if X is a prefix of X$'$.}

When a router receives an interest for name X and there are no
pending interests for the same name in its PIT (Pending
Interests Table), it forwards the interest to the next hop, according
to its routing table. For each forwarded interest, a router stores
some state information, including the name in the interest and the
interface on which it was received. However, if an interest for
X arrives while there is already an entry for the same name in the
PIT, the router collapses the present interest (and any subsequent
ones for X) storing only the interface on which it was received. When
content is returned, the router forwards it out on all interfaces from which
an interest for X has been received and flushes the corresponding
PIT entry. Note that, since no additional information is needed to deliver
content, an interest does not carry a ``source'' address. Any NDN router can 
provide content caching; its magnitude is limited only by resource availability. 
Consequently, content might be fetched from routers caches, 
rather than from its original producer. (Hence, no ``destination'' addresses 
are used in NDN).
Further details of NDN architecture can be found in \cite{Jacobson2009}.

NDN deals with content authenticity and integrity by making
digital signatures mandatory for all content. A signature
binds content with its name, and provides origin authentication no
matter how, when or from where it is retrieved.
Public keys are treated as regular
content: since all content is signed, each public key content
is effectively a ``certificate''. NDN does not mandate any
particular certification infrastructure, relegating trust management
to individual applications.
Private or restricted content in NDN is protected via encryption by the
content publisher. \begin{figure}[tb]
\center
\includegraphics[width=2.7in]{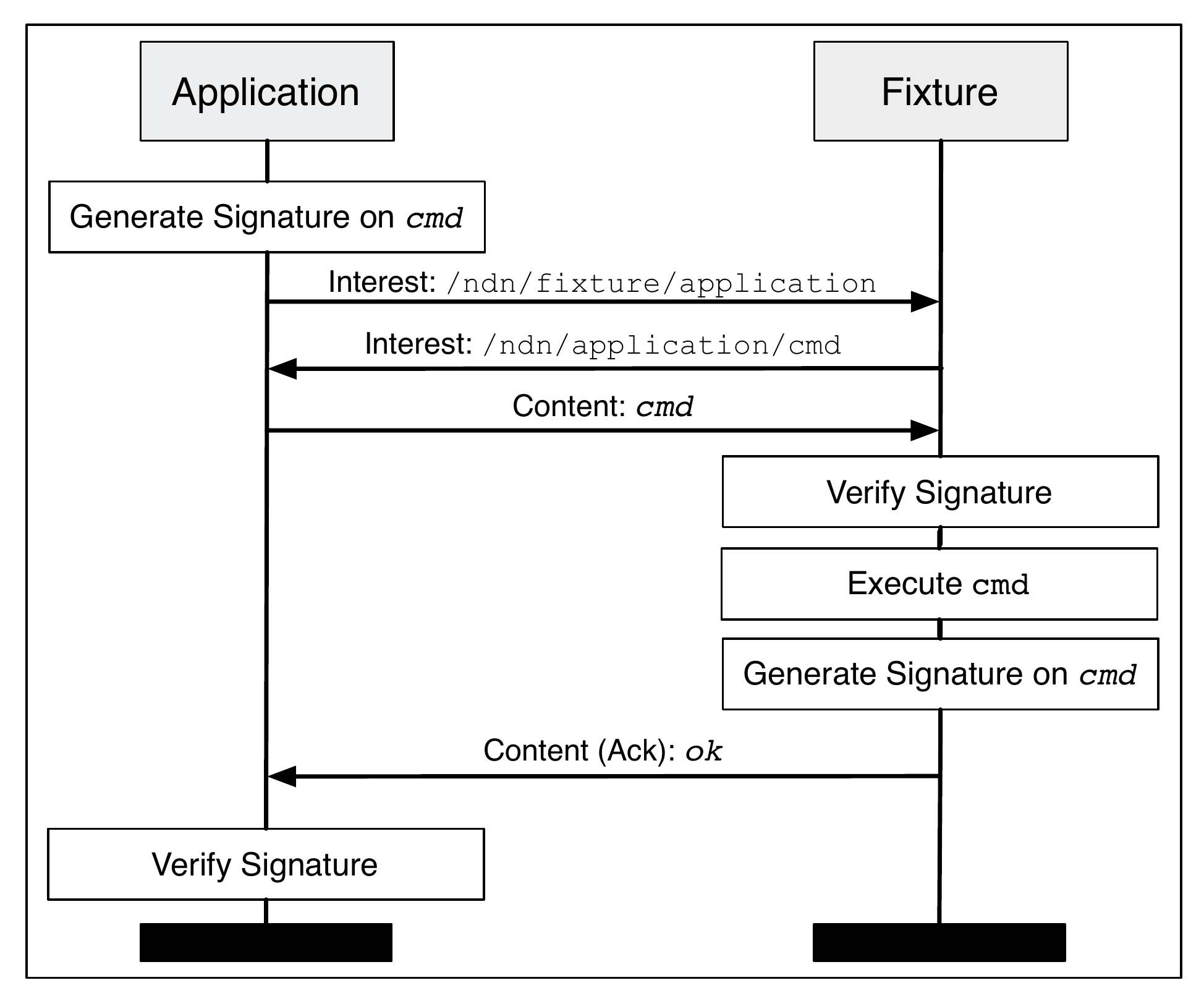}

\vspace{-0.35cm}
\caption{Base-line protocol.}
\label{fig:basic-proto}
\end{figure}

\begin{figure}[t]
\center
\includegraphics[width=2.7in]{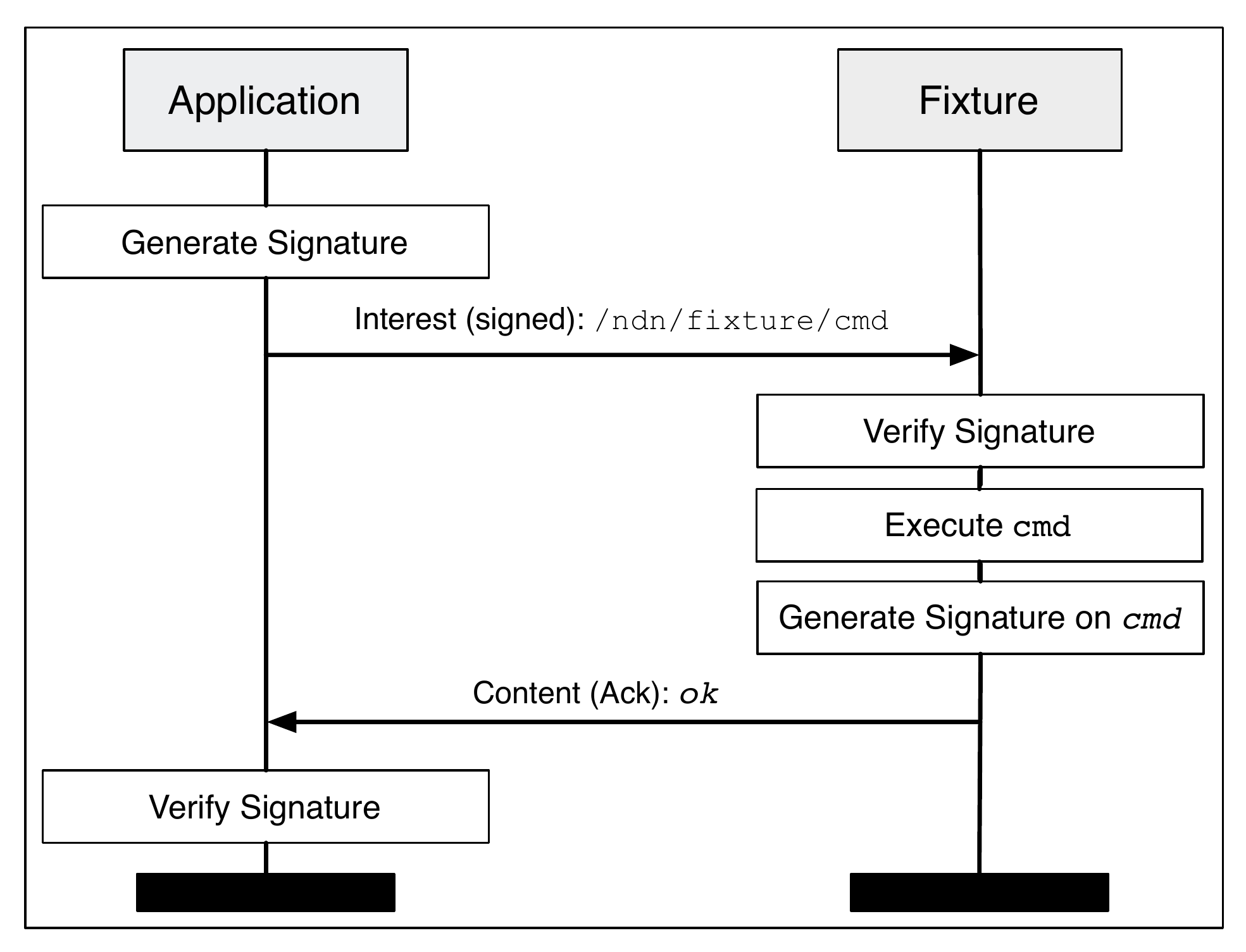}

\vspace{-0.35cm}
\caption{Protocol with authenticated interests.}
\label{fig:auth-int-proto}
\end{figure}

\section{Lighting Control over NDN}
\label{sec:build-autom}
Our setup involves four parties: a configuration manager (\CM), one or more fixtures (\Fixt), 
one or more applications (\App) and an authorization manager (\AM). \CM\ is in charge of the 
initial fixture configuration. This includes, on a per-fixture basis: assigning a fixture its NDN 
namespace, installing a trusted public key (owned by \AM) that identifies the local domain, 
and giving a fixture its identity represented by a unique public key. 
Note that, in NDN parlance, ``namespace'' refers to content published by some entity, whereas,
``identity'' refers to a public key associated with some entity that publishes content.
\AM\ determines which applications are allowed to access 
each fixture, signs applications' public keys and (optionally) issues signed access control lists.
While \CM\ and \AM\ represent distinct functions, in practice, they are likely to be physically 
co-located.

\subsection{Base-Line Protocol}
We start by observing that NDN can be easily used to securely implement basic lighting control 
without requiring any new features or components. As shown in Figure \ref{fig:basic-proto}, when 
application \App\ needs to send a command to fixture \Fixt, the base-line protocol works as follows: 
\begin{compactenum}
\item \App\ creates (and signs) a new content object $cmd$ containing the desired command. 
\item \App\ issues an interest $int_A$ with a name in \Fixt's namespace that references the name of $cmd$.
\item \Fixt\ receives $int_A$, stores it in its PIT, and issues an interest $int_F$ for the name of $cmd$.
\item \App\ receives $int_F$ and responds with $cmd$. 
\item \Fixt\ receives $cmd$, (1) checks its access control list to determine if \App\ is authorized to 
execute the command in $cmd$, (2) verifies the signature of $cmd$, (3) executes the command, and (4) 
replies with an acknowledgement (from here on abbreviated as ``ack'') in the form of a new signed 
content object $ack$. Finally, \Fixt\ flushes $int_A$ from its PIT.
\item \App\ receives $ack$ and verifies its signature.
\end{compactenum}
The main drawback of this protocol is its high latency and bandwidth overhead: a single 
command requires 4 rounds and 4 messages, instead of the ideal 2 rounds/messages 
(Also, as is well-known, protocol robustness suffers and complexity increases with the number of rounds.)
Thus, this approach is a poor match for delay-sensitive lighting control.

Alternatively, \Fixt\ could continuously issue interests that solicit \App's commands. This way, whenever 
\App\ issues a new command, it does so by simply satisfying the most recent interest. While this approach 
would address the latency issue of the base-line protocol, it introduces new problems. 

First, \Fixt\ would have to always issue one interest per each \App\  allowed to control it. In an installation
with multiple applications ($m$) interacting with a large number of fixtures ($n$), the overhead of periodic
$O(mn)$ interests would be significant. Also, an application would be unable to generate a rapid burst of 
commands to the same fixture, since the latter would operate in a lock-step fashion. (In other words, \App\
can only issue a new command after it receives an interest from \Fixt).

\subsection{Whither Authenticated Interests?}
We now consider another approach that, at least in principle, violates the tenets of NDN. Recall that NDN stipulates that
all content objects must be signed. Each entity implementing the NDN protocol stack must be able to 
verify content signatures. Interests, however, are not subject to the same requirement. One reason 
for this design choice is efficiency: public-key signature generation and verification is expensive. Moreover, signatures from different parties prevent straightforward interest aggregation.
Another reason is privacy: traditional public-key signatures carry information about the signer.
There are, however, applications that could benefit from authenticated interests and control of 
building systems like lighting 
seems to be one.

Authenticated interests can be implemented using both public key and
symmetric authentication mechanisms,  i.e., signatures and message authentication codes (MACs), 
respectively. For the sake of generality, we refer to the output of both as {\em authentication tags}.
Due to the flexibility of NDN naming, where name components can be application-determined 
and are opaque to the network, an authentication tag can be placed into an NDN name as a
{\em bona fide} component thereof. This way, an authentication tag becomes transparent to NDN routers
and only the target of the command 
 would interpret and verify it.

While MACs obviously perform much better than signatures, they make auditing difficult: a MAC cannot 
be attributed to one party. For this reason, if there is an auditing requirement -- and if timing constraints allow -- we 
prefer signature when fixtures are used as actuators (i.e., when commands generate feedback). \linebreak Whereas,  
MACs are naturally preferred when fixtures act as sensors, i.e., when applications retrieve information from them.
Regardless of their type, computation of authentication tags must be randomized to ensure uniqueness, based on
either nonces or timestamps. However, this would inhibit  aggregation of authenticated interests, since each authentication tag 
would be distinct; hence, no two names would be the same, with overwhelming probability.

In general, using authenticated interests is fairly straightforward, as illustrated in Figure \ref{fig:auth-int-proto}. \CM\ configures each 
fixture with a specific namespace and \AM\ assigns a set of rights to each application, tied with the application public key 
or to a symmetric key shared with the fixture.  The name reflected in an authenticated
interest would contain three parts: (1)  prefix part (used for routing) that corresponds to the fixture namespace,
(2) actual command, and (3) randomizer (nonce or timestamp) along with the authentication tag computed
over the rest of the name: 
\begin{center}
\resizebox{0.465\textwidth}{!}
{
\fbox{
$
\underbrace{\mbox{\tt/ndn/fixture-namespace}}_{(1)} 
\underbrace{\mbox{\tt/command}}_{(2)}
\underbrace{\mbox{\tt/randomizer||auth-tag}}_{(3)}
$
}}
\end{center}
The idea is that, when \Fixt\ receives such an interest, it verifies the authentication tag, executes the command
and replies with a signed ack as content. 
The first task (verifying the authentication tag) is simple only if
one application controls the fixture or if part 1 of the name somehow uniquely identifies the requesting application.
Whereas, if multiple applications are allowed to issue the same class of commands to a given fixture and
use the same type of authentication tags, the fixture would need to determine the exact application
by repeatedly verifying the authentication tag. This could translate into costly delays. 
This issue can be easily remedied by overloading NDN names even further and including another
explicit part that identifies the requesting application (or its key). 
Another drawback of this (and the base-line) approach is that the fixture needs to sign, in real-time, 
the acknowledgement, which is represented as content. Since a typical fixture is a relatively anemic
computing device, signature generation might involve non-negligible delays.

On the other hand, authenticated interests offer much faster (2-message/2-round) operation than the base-line
protocol described in the previous section. Furthermore, an application can issue multiple commands
to the same fixture in rapid succession, i.e., without waiting for an ack. (However, note that this 
could have negative consequences if closely-spaced interests arrive out of order).

\subsection{Secure Lighting Control Framework}
\label{sec:framework}
Based on the preceding discussion, we conclude that a more specialized approach to secure lighting control 
over NDN is necessary in order to obtain reasonable performance while adhering to NDN principles.
To this end, we construct a security framework that includes:
\begin{compactitem}
\item 
A trust model wherein public keys are associated with NDN namespaces. The framework relies on this functionality to determine the entity 
that ``owns'' a particular namespace. For example, this allows us to ensure that a content object issued by a fixture in response to a 
command has been generated by the correct party.
\item
A syntax for key attributes and access control policies that binds a public key with its attributes, as determined by the signer (certifier) 
of this key.
\item 
A protocol that defines how fixtures are initialized and how applications and fixtures handle authenticated commands.
\end{compactitem}
In the design of our framework, we consider an adversary that can control the communication channel between \App\ and \Fixt, i.e., it can record, drop, modify, inject, delay, or replay any packet. 
The goal of the adversary is to (1) produce a command of its choice that is successfully executed by \Fixt; or (2) undetectably delay, or replay legitimate commands from \App; or (3) provide an acknowledgment to \App\ for a command that has not been executed.

\subsection{Trust Model} \label{sec:trust-model} 
NDN does not mandate the use of any particular trust model: each application is free to adopt the trust model
that best suits it.  Our trust model allows an entity (e.g., applications and fixtures) to publish content only in 
its namespace or any of its children. ($name_A$ is a child of $name_B$ if the latter is a prefix of the former).

Our trust model implements this restriction using public-key cryptography. 
Zero or more public keys are associated with each namespace. A content object published under namespace 
$name$ must be signed using the key associated to $name$ or any of its ancestors. 

A trusted third party (TTP) -- e.g., \AM\ -- generates the key-pair $K_{root}=(pk_{root},sk_{root})$ and distributes 
$pk_{root}$. This public key is used as root of trust; a signature on a content object computed using $sk_{root}$ is 
always accepted. In order to associate $pk_{P}$, belonging to producer $P$,  with namespace $name_P$, 
TTP publishes, under $name_P${\small \tt/key}, a content object containing $pk_{P}$. 

$P$ can delegate a key $sk'_{P}\neq sk_{P}$ to sign content in namespace $name_P${\small \tt/sub-namespace} 
by publishing the corresponding public key $pk'_{P}$ under $name_P${\small \tt/sub-namespace/}{\small \tt key}. 
This mechanism allows TTP to delegate some of its certification capabilities to each producer.

$P$ can prove to anyone its ownership of a key linked to $name_P$ through a simple 
challenge-response protocol.  The challenger issues an interest for a content object with name 
$name_P${\small \tt/nonce} where {\small \tt nonce} 
is a fresh random string selected by the challenger. $P$ is able to respond with a valid content object only if it owns the 
signing key linked to $name_P$, one of its ancestors, or TTP's signing key.

While our implementation is based on RSA signatures, hierarchical identity-based signature (HIBS) 
\cite{DBLP:conf/asiacrypt/GentryS02} schemes represent a viable alternative. A HIBS scheme is a signature scheme 
where any string can be a public (i.e., verification) key. Private (signing) keys are generated by a key generation center. 
Given a signing key $sk$ corresponding to a string $s$, $sk$ is also a valid signing key for any string $s||t$ where ``$||$'' 
denotes string concatenation. Moreover, $sk$ can be used to compute a new signing key $sk'$ corresponding to 
$s||u$ for any string $u$.

With HIBS, the public key corresponding to a namespace $name$ is the string representing $name$. TTP acts as a key generation center and issues signing keys to producers, each key corresponding to a namespace.

The main drawback of an implementation based on HIBS is the lack of support from the current 
NDN codebase. As such, routers cannot verify HIBS signatures on content objects. 
For this reason, our lighting control prototype relies on RSA signatures.

\subsection{Key Attributes \& Access Control Policies} \label{sec:key-attributes} 
All attributes of a public key are expressed using the name under which such key is published. 
Each attribute is a name/value pair expressed as two consecutive namespaces: the first indicates a key attribute name, 
and the second -- its value.

Recall that an NDN content object is bound to its name by a public-key signature. According to our trust model, 
such a signature must be issued either by the TTP or by the owner of the namespace that contains the public key.
The set of attributes defined in our framework is detailed in table \ref{tab:attributes}.
\begin{table}[htdp]
\small
\begin{center}
\begin{tabular}{|l|l|}
\hline
{\bf Attribute} & {\bf Description} \\
\hline
{ \tt domain} & application's domain\\
\hline
{ \tt appname} & application identifier\\
\hline
{ \tt access} & application's permissions on the fixture\\
\hline
{ \tt expires} & expiration date in \\
& {\em generalized time} notation: \\
& (YYYYMMDDHHMMSSZ)\\
\hline
\end{tabular}
\caption{Attributes}
\vspace{-0.5cm}
\label{tab:attributes}
\end{center}
\end{table}%
Applications can extend this set with new attributes. For example, a public key $pk_P$ published under 
{\small \tt /ndn/uci/ics/432B/domain/lighting-domain-1/\\appname/light-board-1/access/full-access/\\expires/20151231235959Z/key} 
specifies that $pk_P$ belongs to application {\em light-board-1} in domain {\em lighting-domain-1}, that has ``full access'' to fixtures in 
such domain. $pk_P$ -- when published under such name -- is considered invalid after December 31, 2015. 

An attribute name can appear more than once with different values. The combined attribute value is computed as the intersection of 
all the instances of such attribute. As with any NDN data packet, the issuer of content object with payload $pk_P$ is specified in the 
content object's {\em key locator} field.

\subsection{The Protocol}
We now introduce the protocol for controlling NDN-connected light fixtures.
The protocol is composed of three sub-protocols:  bootstrapping, application authorization and control.

\medskip\noindent{\bf Bootstrap. } 
New fixtures must be {\em paired} with \CM\ and {\em bootstrapped} in order to be able to receive commands from 
applications. The pairing process consists of the distribution of a (short) symmetric key from \Fixt\ to \CM. For example, in our implementation 
\CM\ scans a barcode on \Fixt's enclosure, that represents a symmetric key factory-installed on \Fixt. 
Next, \CM\ initializes \Fixt. 

Fixture initialization consists of selecting an NDN name for \Fixt, (loosely) synchronizing \CM\ and \Fixt\ clocks and 
installing (on \Fixt) a trusted public key that belongs to \AM. This public key identifies the domain under which \Fixt\ operates.
\CM\ then communicates a signing key-pair to \Fixt.\footnote{\scriptsize Alternatively, \Fixt\ can generate a signing key-pair and communicate 
the resulting public key to \CM.} This key-pair is linked to \Fixt's namespace, and it represents the identity of \Fixt. 
Additionally, \CM\ can specify the NDN name of one or more ACLs that \Fixt\ must use to determine applications' permissions. 
At this time, \Fixt\ also generates a long-term secret master key $k_\Fixt$. This key is optionally used later to derive application-specific symmetric keys (i.e., $k_{\App}$) for authentication purposes.
Once \Fixt\  is correctly initialized, it responds with the current time and a hash of all the information exchanged during bootstrap. 

\medskip\noindent{\bf Application Authorization. }
\AM\ grants control privileges to an application by signing the latter's public key. Given 
$pk_\App$ belonging to $\App$ and intended permissions $perm_\App$,  \AM\ first
constructs a namespace $name_\App$ containing ``{\small \tt access/}$perm_\App$'', 
as specified in Section \ref{sec:key-attributes}. Then it  signs $pk_\App$ and publishes it (as content) 
under $name_\App$.
Any fixture under control of \AM\ can verify that $\App$ owns permission $perm_\App$ by 
asking it to prove ownership of namespace $name_\App$, as in Section \ref{sec:trust-model}.

\medskip\noindent{\bf Control Protocol. }
The protocol is designed for resource-constrained fixtures interacting with a large number of applications. Thus, we aim to 
minimize computation and communication costs and amount of memory required to perform interest authentication.
We avoid storing per-application long-term information (e.g. application keys) on each fixture. A fixtures stores a constant 
amount of  state for each application currently interacting with it. We emphasize that, in order to issue and verify commands, 
applications and fixtures do not need to communicate with either \CM\ or \AM.

Application \App, that owns a key distributed under $name_\App${\small \tt /key}, issues an interest with command 
{\small \tt cmd} for fixture \Fixt\ with NDN name $name_\Fixt$ as follows:
\begin{center}
\fbox{$name_\Fixt${\small \tt/}$name_\App${\small \tt/cmd/auth-token}}
\end{center}
The string ``{\small \tt cmd}'' represents a fixture-specific command. Since our framework does not specify any particular 
format for commands, this string is simply treated as an {\em opaque} binary field.
For example, a simple command could be: ``{\small \tt on}'' or ``{\small \tt off}'', while a more complex one could be:
``{\small \tt intensity/+10/rgb-8bit-color/F0FF39}''. 

The field {\small \tt auth-token} encodes the command authentication token, constructed as:
{\small \tt state} $||$ {\small \tt authenticator}. 
The first part represents state information required to prevent timing and replay attacks. It is, in turn, composed of:
sequence number, timestamp and estimated round-trip time (RTT) between \App\ and \Fixt.
The {\small \tt authenticator} part is a signature or a MAC. In either case, it is computed over: 
``$name_\Fixt${\small \tt/}$name_\App${\small \tt/cmd/state}''.
\App\ signs its commands using the private counterpart of $name_\App${\small \tt /key}. 
The key used to compute and verify commands 
authenticated with MAC is negotiated between \App\ and \Fixt\ as detailed below.

When a fixture receives an interest $name_\Fixt${\small \tt/}$name_\App${\small \tt/cmd/}\linebreak {\small \tt auth-token}, 
it determines whether to execute {\small \tt cmd}, as follows:
\begin{compactenum}
\item Verifies that {\small \tt cmd} is well formed.
\item Examines attributes in $name_\App$ to determine whether $\App$ is allowed to issue {\small \tt cmd} (e.g., \Fixt\ check whether $name_\App$ expiration date and access fields). 
\item If available, uses a local or remote ACL specified by \CM\ during the bootstrap phase.
\item Verifies the state of the command. First determines whether the interest is current (also using the estimated RTT 
value as additional information). Then, if it has is no record of previous commands from \App, \Fixt\ extracts the 
sequence number from {\small \tt auth-token} and stores it as: $(name_\App, sequence\ number)$. Otherwise, it checks that the 
stored sequence number is lower than the one in {\small \tt auth-token}.
\item Verifies {\small \tt authenticator} -- signature or MAC on the interest. In case of signature, \Fixt\ retrieves public key $name_\App${\small \tt /key} and stores it in its local cache.
\end{compactenum}
If a pair $(name_\App, sequence\ number)$ stored by \Fixt\ is not updated for a predetermined amount of 
time,  it is considered stale and deleted. This way, at any given time, a fixture only retains state information related 
to active applications.

\begin{table}[htbp]
\centering
\small
\begin{tabular}{|c|l|}
\hline
$\ENC_k(\cdot)$ & Symmetric encryption algorithm \\ & (e.g. AES)\\
\hline
$\HASH(\cdot)$ & A collision-resistant hash function \\ & (e.g. SHA-256)\\
\hline
MAC$_k(\cdot)$ & Message authentication code\\
&   (e.g. HMAC-SHA-256)\\
\hline
PRF$_k(\cdot)$ & Pseudorandom function\\
\hline
$name_i$ & NDN name associated with entity $i$\\
\hline
\end{tabular}
\caption{Notation}
\vspace{-0.3cm}
\label{tab:notation}
\end{table}

\medskip\noindent{\bf Symmetric Authentication. } \label{sec:symm-auth} 
By default, fixtures and applications authenticate commands and acks using public-key signatures. 
However, for performance reasons, they can switch to MAC-s at anytime, which requires establishing a 
shared secret key.  Recall that, at bootstrap, \Fixt\ generates a long-term secret key $k_\Fixt$. 
When \App\  asks \Fixt\ to switch to symmetric authentication, the latter uses $k_\Fixt$ to compute an application-specific
key $k_\App$. After verifying that \App\ owns the namespace $name_\App$ (see Section \ref{sec:trust-model}), \Fixt\ 
computes $k_\App = $ PRF$_{k_\Fixt}(name_\App)$. Then, \Fixt\ sends $k_\App$ to \App\ encrypted under public 
key $name_\App${\small \tt /key}. 
Note that \Fixt\ does not need to store these application-specific symmetric keys: it can compute $k_\App$ from $k_\Fixt$ whenever needed.
Therefore, the amount of symmetric-key-related state stored by \Fixt\  does not depend on the number of authorized applications.

\subsection{Command and Ack Privacy}
We now consider privacy of commands and acks. An eavesdropper may want to learn various parameters in a command or a corresponding ack. 
An application conceals this information by encrypting the command before constructing the name that goes into an 
interest. If symmetric authentication is used, the command encryption key is derived from the MAC key. If the interest is signed, 
the command is encrypted using \Fixt's public-key.
NDN already provides a framework for content encryption \cite{ndn-encryption} that we use to implement 
ack privacy. 

For efficiency reasons, we do not conceal sizes of either commands or acks, thus potentially allowing the
adversary to distinguish among types of commands. Although it is easy enough to 
introduce padding (though incurring costs), more sophisticated attacks exploiting {\em side channels} (e.g., time required for a 
fixture to respond to a command or some other observable feedback) are much more difficult to address. 
Since one the main goals of our approach is generality,  we do not implement any countermeasure for this class of attacks.

\subsection{Ack Authentication}
While not typical today for lighting, we desire that a fixture and other actuators should
provide feedback after processing a command. 
In our NDN context, this naturally 
results in a closed-loop control system and allows NDN routers to flush PIT entries corresponding to processed commands (interests).
For obvious security reasons, acks must be authenticated. (NDN anyhow requires all content to be 
signed). However, in resource-constrained environment of light fixtures, the cost of computing per command (or per ack)
public-key signatures is quite high, especially, considering that a fixture might receive numerous closely-spaced commands.
For this reason, we propose an NDN extension allowing fixtures to efficiently produce authenticated command acks.

A natural and efficient alternative to public-key signatures are symmetric MACs. An application and a fixture could share a key,
and use to authenticate acks, i.e., replace a signature on the content object (that carries the ack) 
with a MAC. Unfortunately, this approach is unworkable if  fixtures and applications communicate through a public network,
specifically, if any NDN routers are involved in \Fixt-\App\ communication. Since MACs are not publicly verifiable, an intervening
NDN router has no means of authenticating MAC-d content and may simply drop it.

Nonetheless, we view NDN as work-in-progress. Thus, we consider end-to-end MAC-based symmetric authentication 
of content as an alternative to publicly verifiable signatures and include it in the implementation.

Next, we describe two techniques that allow public verifiability of acks without requiring public-key operations by 
fixtures,  applications or NDN routers.

\smallskip\noindent{\bf Encryption-based Authentication. }
This technique assumes that \App\ and \Fixt\ share a symmetric key $k$, itself derived from \Fixt-\App\ shared key 
$k_\App$, which is generated at bootstrap time.
To begin, \App\ generates a random $s$-bit value $x$ and, using a block cipher $\ENC$ with block size $s$, 
computes $y=\ENC_k(x)$, $z=\HASH(x)$ where $\ENC$ is used in the ECB mode.
\App\ includes the pair $(z,y)$ as part of the command to \Fixt. Recall that this command is 
represented as an NDN interest
and, on the path to \Fixt, it leaves state in all intervening NDN routers.

Having received an interest, \Fixt\ extracts $x'$ from $y$ as $x'=\ENC^{-1}_k(y)$ and re-computes $z'=\HASH(x')$.
If $z' \neq z$, then \Fixt\ aborts; otherwise, it issues an ack in the form of an empty content object with 
$x$ as a signature. 

Although $x$ is clearly not an actual signature, this technique allows public verifiability. An NDN router 
that observes 
the (ack) content object carrying $x$ must have a corresponding interest (and therefore $z$) in its PIT.
It can efficiently determine the relationship between the interest and the content  by
checking whether $\HASH(x)\stackrel{?}{=}z$.

Commands that are not acknowledged can be retransmitted until they time out. Once a command expires, it must be reissued 
using a new challenge. Despite public verifiability, \App\ cannot prove to a third party that it successfully interacted with \Fixt.
This is because \App\ can unilaterally produce any number of challenge/response pairs without any interaction with \Fixt. 
This motivates a stronger (hash-chain-based) technique described below.

\smallskip\noindent{\bf Hash-Chain-based Authentication. }
The present technique allows \App\ to prove to any third party that it successfully interacted with \Fixt,
with no need for any shared keys; in particular, \Fixt-\App\ interactions become auditable.
This  is very useful in certain circumstances. For example, consider an emergency lighting system: an emergency 
response team can issue an ``alarm'' command to each fixture and turn on all lights.
In case of post-incident investigation, the emergency response team can prove that it issued required commands 
by producing acks from the appropriate fixtures. Another example is a building security 
system, that, in case of an alarm, needs to trigger security lights; proving that lights were turned on correctly 
can be crucial for subsequent evidence gathering.

However, due to its lock-step feature (see below), this technique is designed for infrequent use. At the same time, it is appropriate whenever traditional signature computation by fixtures is too costly -- i.e., when the framework is instantiated on resource-constrained devices that cannot perform public-key operations. 

 Hash-chain signature schemes are particularly
appealing for low-powered devices due to the reduced resource burden of traditional signature schemes,
which have lead to several protocols incorporating their use \cite{asokan, DingMT02, ChallalBH04, YavuzN09}.
Our hash-chain-based authentication method is somewhat similar to Server-Supported Signatures (S$^3$) 
scheme \cite{asokan}. However,  unlike S$^3$, it does not rely on the server and 
does not involve the use of public key cryptography in the generation or verification of ack-s. 
However, it also does not provide all functionality offered by S$^3$; in particular, our method 
can not be used to authenticate an arbitrary payload.

We denote recursive application of hash function $\HASH(\cdot)$ $i$ times to input 
$x$ as $\HASH^i(x)$, i.e., $\HASH^0(x) = x$ and $\HASH^i(x) = \HASH^{i-1}(\HASH(x))$.
A {\em hash chain} is a sequence $\{x, \HASH^1(x),\ldots,\HASH^\ell(x)\}$ for some secret $x$
and $\ell>0$. The last link in the chain $\HASH^\ell(x)$ is called an {\em anchor}.
Hash chains are commonly used for authentication as follows: Alice selects
a random secret value $x$ and sends $\HASH^\ell(x)$ to Bob through an 
authenticated channel. Bob authenticates Alice by challenging her with the value
he currently stores -- $\HASH^i(x)$ for some $i\leq~\ell$. Alice responds
with a pre-image $\HASH^{i-1}(x)$ of $\HASH^i(x)$. Since $\HASH(\cdot)$ is assumed to be 
collision-resistant, knowledge of $\HASH^{j}(x)$ with $j\leq i$ is required to respond to the challenge.

In our protocol, \Fixt\ generates a secret seed $x$
for a hash chain $C$ of length $\ell$. The anchor $\HASH^\ell(x)$, along with other
parameters (including $\ell$) is signed by \AM\ or by \Fixt\ itself, with its own private key.
The result is essentially a certificate valid for up to $\ell$ signatures.
\Fixt\ sends this certificate to \App, either off-line or on-line, whenever it depletes the previous
hash chain. 
For each command, \App\ includes the last value $\HASH^i(x)$
received from \Fixt. \Fixt\ responds to a command with an ack
containing: $\sigma$, $\HASH^\ell(x)$ and $\HASH^{i-1}(x)$.
\App\ can then prove to a third party that it successfully interacted 
with \Fixt\ at least $\ell-i$ times by revealing anchor $\HASH^\ell(x)$, 
the certificate (for the anchor) and the last preimage $\HASH^i(x)$ received from \Fixt. 

The main drawback is that \App\ cannot issue a new command 
until it receives an ack from \Fixt\ for its last command. 
This lock-step approach makes the technique unsuitable if \App\ needs to 
have multiple commands {\em in flight} for any given fixture. 
Moreover, \Fixt\ must use a different hash chain for every controlling  application.

\smallskip\noindent{\bf Packet Loss. }
Either the interest from \App\ to \Fixt\ or the corresponding ack might be lost.
Clearly, \App\ cannot distinguish between the two cases. After issuing a command, if the 
ack is not received, \App\ continues to issue {\em the same} interest for a 
predetermined amount of time. If still no ack is received , the command
is aborted and \App\ and \Fixt\ fall back to authenticating acks through 
regular public-key signatures.
\Fixt\ can easily determine if a received interest corresponds to a retransmission or to a 
new command by checking whether the preimage of the challenge $\HASH^i(x)$ 
in the interest has already been revealed. If the command is a re-transmission and 
the original has already been acknowledged, \Fixt\ simply issues a new 
ack containing the pre-image of $\HASH^i(x)$; this requires no computational effort.

\section{Prototype Evaluation}
\label{sec:performance}
In order to evaluate the performance of the proposed architecture, we implemented a library -- 
called {\tt NameCrypt} -- designed for lighting control in a theatrical environment. We also deployed it
in an actual theatrical lighting installation. In this setting, applications and lighting fixtures 
interact over a local-area network. 

Our setup involves three applications: (1) a sequencer that outputs pre-generated patterns, (2) 
a controller that uses algorithmic patterns and (3) a fader. 
These applications control eleven 
lights, connected to five embedded devices (described below). The embedded devices, in turn, are connected to the lights using  
KiNet~\`{kinet}, a proprietary Philips protocol which runs on TCP/IP over ethernet.

\begin{table}[tdp]
\begin{center}\resizebox{0.4\textwidth}{!}
{

\begin{tabular}{|l|c|c|}
\hline
Operation & Intel           & ARM \\
                 & Core2Duo & Cortex A8 \\\hline\cline{1-3}
Create auth. command (RSA-1024) & 1.981 ms & 21.553 ms \\\hline
Verify command (RSA-1024)& 0.096 ms & 0.435 ms \\\hline
Compute HMAC key from \Fixt's secret & 0.005 ms & 0.046 ms\\\hline
Create auth. command (HMAC) & 0.006 ms & 0.067 ms \\\hline
Verify command (HMAC) & 0.013 ms & 0.152 ms \\\hline
\end{tabular}}
\end{center}
\vspace{-0.5cm}
\caption{Performance of RSA and HMAC authenticated commands on Intel and ARM platforms.}
\label{tab:perf-interests}
\end{table}%

\subsection{Implementation}
{\tt NameCrypt} provides low-level functionality required by our protocol on top of CCNx. 
It is implemented in C interfacing to OpenSSL for cryptographic services and to CCNx 
for transport. 
The target platform of the lighting fixture is an off-the-shelf embedded device based on the
Gumstix Overo Air~\cite{gumstix} computer-on-a-module.  This device is running a 600 MHz 
Texas Instruments OMAP 3503 ARM Cortex-A8 CPU with 256MB RAM. It supports both 
WiFi and Fast Ethernet and runs Linux kernel 3.0. In our tests, we used CCNx version 0.4.0.
Our code is portable and has been also successfully deployed on both Intel-based Mac and 
Linux computers.

Interests are signed using RSA with a 1024-bit modulus and a public exponent of 3. 
This allows for efficient verification, requiring only requires two multiplications. 
The hash algorithm used in signature computation is SHA-256.
We implemented symmetric authenticated interests using HMAC with SHA-256.

Each command is associated with a state variable,
which contains current time in seconds/microseconds, sequence number corresponding to the 
command and an estimated round-trip time between application and light fixture expressed in 
milliseconds. 

Commands are treated as opaque binary strings by {\tt NameCrypt}. Since CCNx allows name components 
to be binary blobs, commands and authenticators are not encoded in any printable format.

{\tt NameCrypt} also implements efficient authentication for fixture acks. Encrypted challenges are 
implemented using AES-128 as block cipher and SHA-256 as the hash function. Hash-chain-based 
authenticated acknowledgments are based on SHA-256.  Our implementation uses a very simple 
``pebbling'' technique to reduce the cost of  acknowledgment generation: rather than storing the 
whole chain (i.e, performing a full lookup) or computing pre-image $\HASH^i(x)$ from $x$, fixtures 
store every 100-th link in the chain. This way, returning $\HASH^i(x)$ requires, on average, 
computing 50 hashes. We emphasize that there are more efficient pebbling technique 
(e.g., ~\cite{Hu05efficientconstructions}) that we can adopt without any changes to our protocol.

\subsection{Performance Evaluation}
Experiments were performed on a commodity laptop, which runs the sequencer, controller and fader (see Section \ref{sec:performance}),  
and on a low-powered embedded system (representative of a lighting fixture or a low-power fixture controller). 
The laptop uses a 2.53GHz Intel Core2Duo CPU. The embedded system uses an the Overo 
platform detailed above. We evaluated command and ack 
authentication micro-benchmarks and discuss performance considerations. 

Table \ref{tab:perf-interests} shows the results of micro-benchmarks in command authentication.
Time required to generate an RSA signature on the Intel platform is comparable to typical network latency 
in a LAN and does not significantly affect the performance of the whole protocol. Verification is well 
below typical network latency on both platforms. For this reason, we believe that features provided by 
digital signatures and their relative low cost justifies their use in an environment where commands are 
generated on non-constrained device. On the other hand, benchmarks show that low-power devices 
are not well-suited for generating real-time signatures on commands. In this case, we recommend the 
use of HMAC.

Symmetric authentication incurs negligible performance impact. Fixtures must generate a symmetric 
key for each application starting from their secret. This requires far less than a millisecond on our test devices.

Table \ref{tab:perf} shows timing results of ack authentication mechanisms.
Similar to command authentication, digital signatures do not introduce any significant delay on the 
Intel platform, while signature generation is relatively expensive on the ARM. Whenever viable, 
our tests show HMAC provides adequate performance. 
However when public verifiability is required and standard signatures are too expensive, encrypted 
challenges are an appealing option, as shown in Table \ref{tab:perf}.

Applications and fixtures may want to rely on hash chains for added functionality. In this case, 
benchmarks clearly show that fixtures should use an efficient representation of the hash chain, 
i.e., one that does not force the fixture to re-generate a large portion of the chain from the secret 
seed $x$. With the ``pebbling'' technique, challenge-response requires less than a millisecond.
 All other operations incur very low overhead on both platforms.

\begin{table}[htdp]
\begin{center}\resizebox{0.42\textwidth}{!}
{

\begin{tabular}{|l|c|c|}
\hline
Operation & Intel           & ARM \\
                 & Core2Duo & Cortex A8 \\\hline\cline{1-3}
Sign content object (RSA 1024) & 2.018 ms & 26.418 ms \\\hline
Verify content object (RSA 1024) & 0.046 ms & 1.301 ms \\\hline

Authenticate/verify (HMAC) & 0.006 ms  & 0.070 ms \\\hline\hline

Encrypted challenge -- create & 0.003 ms & 0.043 ms \\\hline
Encrypted challenge -- answer & 0.001 ms & 0.015 ms \\\hline
Encrypted challenge -- verify & 0.001 ms & 0.015 ms \\\hline\hline

Hash chain -- create & 11.350 ms & 88.407 ms\\\hline
Hash chain -- answer w/ lookup & $<$0.001 ms & $<$0.001 ms\\\hline
Hash chain -- answer w/o full lookup & 5.104 ms & 44.196 ms\\ \hline
Hash chain -- answer w/ partitioning & 0.052 ms & 0.443 ms\\ \hline
Hash chain -- verify & 0.001 ms & 0.010 ms \\\hline
\end{tabular}}
\end{center}
\caption{Performance analysis of ack authentication. 
RSA with public exponent $3$; hash chains 
with $10,000$ elements}
\label{tab:perf}
\end{table}%
\section{Security Analysis}
\label{sec:security}

\medskip\noindent{\bf Trust model.} 
In our model the relationship between content objects can be represented in a undirected graph where each content object is a vertex. Various vertices are connected through public-key signatures. In particular, each edge represents a pair (public-key signature, key locator) -- i.e., a signature is an edge from a content object carrying a public key to a content object signed by that key, while a key locator is an edge from a content object to the content object carrying the corresponding verification key.
Vertices carrying public key can have multiple edges, while vertices representing regular content have only one edge.

A proof of ownership of $name_P$ from producer $P$ to a challenger $C$ consists of a graph with a path from a vertex $V_i$, which represents the content object named by $C$ under $name_P$, to vertex $V_0$, which denotes the content object containing TTP's key. The prefix of the NDN name of each vertex along the path from $V_0$ to $V_i$ must be the full namespace of the previous vertex, with the exception of the vertex signed by the TTP.
For this reason, only the TTP or the owner of a namespace $name$ can elect a user to be the owner of namespace $name${\small \tt/}$child$.

\medskip\noindent{\bf Key Attribute and Access control policies. }
Let application \App\ be the owner of a namespace $name_\App$, which represents a set of pairs attribute/value as defined in Section \ref{sec:key-attributes}. 
The goal of our key attribute and access control policy mechanism is to guarantee that -- without the ownership of additional namespaces -- \App\ can only assign new namespaces to other owners as long as such namespaces identify more restrictive attribute/value pairs (as defined by specific applications) than $name_\App$.

The security of our key attribute and access control policy mechanism is based on the security of our trust model. In particular, our trust model guarantees that \App\ cannot become owner of a namespace with a prefix that differs from the full name of $name_\App$ without receiving any additional namespace ownership from TTP or other applications. This is implied by the well-formedness of the proof of ownership: each vertex on the path from the TTP's key to \App's key must represent a namespace with a prefix corresponding to the full namespace of the previous vertex -- with the exception of the first vertex after TTP's key. A proof that shows \App's ownership of a namespace with a prefix that does not correspond to the full name of the parent namespace would clearly be invalid.

\medskip\noindent{\bf Symmetric authentication mechanism. }
We now analyze the security of our symmetric interest authentication mechanism as defined in Section \ref{sec:symm-auth}. 
Consider a malicious application \Adv\ whose goal is to issue correctly authenticated commands for 
\Fixt\ in a namespace for which \Adv\ has never received the corresponding symmetric key from \Fixt.
We model this adversary as follows: \Adv\ is allowed to query \Fixt\ with any arbitrary namespace $name_i$ and receive the corresponding key $k_i = $ PRF$_{k_\Fixt}(name_i)$. 
Eventually \Adv\ selects a namespace $n_{adv}$, never queried to \Fixt, under which it wants to be challenged. After revealing $n_{adv}$, \Adv\ is allowed access to oracle $O^{(n_{adv})}(cmd_i)$, which outputs $m_i = \mbox{MAC}_{k_{adv}}(cmd_i)$ where $k_{adv}=\mbox{PRF}_{k_\Fixt}(n_{adv})$. $n_{adv}$ cannot be included in subsequent queries to \Fixt. 
The goal of \Adv\ is to issue a pair $(cmd_{adv}, m_{adv})$ with $m_{adv}$ $=$ $\mbox{MAC}_{k_{adv}}$ $(cmd_{adv})$, with $cmd$ never queried before to $O^{(n_{adv})}$. 
In other words, \Adv\  can obtain the symmetric key corresponding to any namespace of it's choice from \Fixt. 
Also, \Adv\ can choose a ``target namespace'', $n_{adv}$, and request commands to be authenticated using $O^{(n_{adv})}$. \Adv's goal is to issue a {\em never requested} authenticated command in namespace $n_{adv}$.

We now sketch how to construct a simulator $S$ that interacts with challenger $C$ for a secure MAC with key $k$ and uses \Adv\ as a subroutine. For each query $n_i$  from \Adv\ to \Fixt\ where $n_i$ was never asked before, $S$ responds with a random value $k_i$. $S$ stores pairs $(n_i, k_i)$ in a table that is used to respond consistently to subsequent queries.
Assuming that the PRF used by \Fixt\ is secure, \Adv\ cannot distinguish between 
 (truly) random responses from $S$ and the pseudorandom responses from \Fixt.
$S$ implements $O^{(n_{adv})}(cmd_i)$ by querying challenger $C$ on $cmd_i$. $C$ returns $m=\mbox{MAC}_{k}(cmd_i)$ and $S$ forwards it to \Adv. This way, 
$S$ implicitly sets  PRF$_{k_\Fixt}(n_{adv})=k$ without knowing $k$.
Eventually \Adv\ outputs $(cmd_{adv}, m_{adv})$, and $S$ outputs $(cmd_{adv}, m_{adv})$ as its response to $C$.
It is easy to see that $(cmd_{adv}, m_{adv})$ is a well formed pair command/authenticator iff $\mbox{MAC}_k(cmd_{adv})= m_{adv}$. Since we assume that MAC is secure, \Adv\ can only output such pair with negligible probability.

\medskip\noindent{\bf Encryption-based Ack Authentication. }
The goal of our encryption-based ack authentication protocol is to prevent an adversary \Adv\
from acknowledging a command $cmd$ on behalf of -- and without any help from -- \Fixt. We argue that, if $\HASH$ is hard to invert and $\ENC$ is a secure block cipher -- i.e., $\ENC$ implements a pseudorandom permutation -- then \Adv\ has only negligible probability of generating a valid ack for $cmd$.\footnote{\scriptsize Our scheme is also secure if $\ENC$ is any CPA-secure encryption scheme with efficiently sampleable ciphertext space. In this case, the value $r$ in pairs $(r,\HASH(x))$ in the security discussion must be selected from the ciphertext space of $\ENC$.}

We model \Adv\ as follows:it interacts with a challenger by requesting pairs $(y_i,z_i) = (\ENC_k(x_i),\HASH(x_i))$ for a value $x_i$ of its choice. Eventually, \Adv\ receives a challenge $(y,z)$ corresponding to a random $x$ selected by the challenger.  \Adv's goal is to output a value $x'$ such that $\HASH(x')=z$.

We argue that the existence of such adversary -- which outputs a correct $z'$ with non-negligible probability -- would either violate the security of the block cipher or the one-wayness of the hash function. In particular, it would allow the construction of either a distinguisher that sets apart the output of $\ENC$ from a truly random string of the same size, or a simulator $S_{adv}$ that inverts $\HASH$.

First, we observe that \Adv\ cannot distinguish between pairs $(\ENC_k(x),\HASH(x))$ and $(r,\HASH(x))$ where $r$ is a random value of the same size as the block size of $\ENC$. If \Adv\ could distinguish between the two with non-negligible advantage, it could be trivially used to build a distinguisher that sets apart the output of a pseudorandom permutation from a truly random string of the same size.

We define $S_{adv}$ as follows: it receives a challenge $y=\HASH(x)$ for a random $x$ chosen by a challenger $C$.  $S_{adv}$ must compute a value $x'$ such that $\HASH(x')=y$. 
$S_{adv}$ answers \Adv's queries $x_i$ by returning $(r_i, \HASH(x_i))$, where $r_i$ is a random value of appropriate size. It also stores pairs $(x_i, r_i)$ to answer consistently to subsequent queries. 
\Adv\ can only detect the simulation by determining that $r_i$ is not the output of $\ENC$. However, we argued above that this is possible only with negligible probability.

Eventually \Adv\ requests a challenge and $S_{adv}$ responds with $(y, r)$ for a fresh random $r$. \Adv\ can detect the simulation only if there exist an index $i$ such that $x_i=x$, which can happen only with negligible probability.

Finally, \Adv\ outputs $x'$. $S_{adv}$ outputs the same value to answer the challenge from $C$. It is easy to see that $S_{adv}$ successfully inverts $\HASH(\cdot)$ iff \Adv's output is correct. Therefore, \Adv\ can only forge authenticated acks with negligible probability.

\medskip\noindent{\bf Hash-chain-based Ack Authentication. }
We consider an adversary \Adv\ that can control the communication channel between \App\ 
and \Fixt\ -- e.g. it can drop, modify, inject, delay, or replay any packet between the two parties. 
The goal of \Adv\ is to produce an authenticated ack to any command issued 
by \App\ for which \Fixt\ has not yet responded. 
 
In our hash-chain-based protocol, upon completion of an authenticated command \Adv\ from \App\ containing challenge $\HASH^i(x)$, \Fixt\ reveals preimage $p$ in its acknowledgment, such that $\HASH(p)=\HASH^i(x)$, i.e., $p=\HASH^{i-1}(x)$. If \App\ fails to receive an authenticated ack from \Fixt, it reissues $cmd$ until either: 1. \App\ receives $p$; or 2. after a predefined timeout. 
We emphasize that \App\ does not issue a new command $cmd$ until it receives an ack for $cmd$ from \Fixt. 

If \App\ deviates from this behavior, the protocol is insecure: consider the scenario where \App\ issues $cmd$ for which it receives no ack. Then \App\ issues a new command $cmd'\neq c$. Since no preimage was received for $cmd$, \App\ has no other option but to construct $cmd'$ using the same challenge as in $cmd$. Assume that \Fixt\ responded to $cmd$ and \Adv\ dropped the ack after learning $p$. From now on \Adv\ can acknowledge any command issued with the same challenge, which is clearly undesirable. 

Since the same challenge is never used twice, \Adv\ cannot issue an authenticated ack without receiving the corresponding preimage from \Fixt. However, since commands are authenticated, \Fixt\ reveals a preimage only if the command has not been altered by \Adv\ and has been successfully executed. For this reason, \Adv\ can only acknowledge commands successfully executed by \Fixt.

We now show \App\ is able to prove to a third party its successful interaction with \Fixt. In order to do so, \App\ must produce the preimage $p = H^{i-1}(x)$, and an anchor $H^{\ell}(x)$ signed by \AM. We point out that \App\ is unable to produce $p$ unless it receives an authenticated ack from \Fixt\ -- which contains $p$. 

Assume that, given $H^i(x)$, \App\ can produce $p$ without interacting with \Fixt. Since \App\ has no additional information on the hash chain, the only way for \App\ to compute $p$ from $H^i(x)$ is to invert $H(\cdot)$. Therefore \App\ can only succeed in producing a valid proof of interaction with negligible probability.

We emphasize that $p$ does not depend on the actual command it acknowledges. For this reason, a proof of interaction does not suffice to determine which command it corresponds to. However if $A$ is authorized to issue only a command $cmd$ with no parameters, then $p$ successfully shows that \Fixt\ must have acknowledged $cmd$.

\section{Related Work}
\label{sec:relwork}
As discussed in Section \ref{sec:Focus}, we target a hybrid design space that draws from 
both the lighting control and building automation worlds.   Here we summarize related
work in both areas. 

Digital lighting control plays a key role in both building automation and entertainment settings. (In fact,
their use in the latter often leads the way in technical innovation \cite{JiangJR10, Huntington}.)
Most current protocols are descended from legacy control architectures based on serial communication~\cite{JiangJR10}. These legacy architectures rely on a separate communication infrastructure. In this context, availability, integrity, privacy and authentication are not an issue since outsiders are assumed to have no access to the communication media.

In order to reduce deployment cost and to evolve such architectures to building- and campus-wide installations, vendors have introduced ways to transport legacy protocol data over IP~\cite{artnet}. This provides great flexibility, allowing devices to reuse already-deployed communication infrastructures such as Ethernet, WiFi, or other IP-compatible media. 
In this transition from serial to IP, vendors have rarely implemented additional security measures~\cite{okabe}, likely to avoid increasing development and deployment costs and in order to maximize compatibility. As a consequence, such protocols must be often run over VLANs, VPNs~\cite{nist-bacnet}, IPSec~\cite{okabe} or physically segregated networks.
Physical segregation is often difficult or even impossible when protocols are running over RF. In this case several of them have been show  to be insecure~\cite{okabe}.

Below we introduce the main protocols in use and briefly discuss their security features or lack thereof.

\medskip\noindent{\bf DMX/DMX512. }
Most modern lighting control protocols descend from or implement DMX512~\cite{dmx}, which is an industry standard multidrop serial protocol based on RS-485. 
 In the past, each serial DMX cable provided 512 one-byte control channels updated at a maximum rate of 44Hz. In DMX nomenclature a link addressing 512 devices is considered a {\em universe}. 
Modern lighting control systems encapsulate DMX payload over modern media such as wired/wireless Ethernet \cite{artnet} or RF \cite{dmx-over-rf}. This has resulted in various competing technologies such as Art-Net~\cite{artnet}, ACN~\cite{JiangJR10}, ETCNet/ETCNet2~\cite{etcnet} which bridge lighting systems and allow them to coexist with newer technology and integrate into BAS.

\medskip\noindent{\bf Art-Net. }
Art-Net \cite{artnet} is a proprietary protocol with open specifications for transporting DMX signal over UDP. It uses wireless/wired Ethernet as communication medium; devices self-configure IP addresses based on the hardware MAC address, and use broadcast transmission for communication. 
Art-Net does not offer any form of authentication or encryption.  Reliance on broadcast and lack of security clearly shows that Art-Net was designed to run on local (access restricted) networks. In \cite{newton05}, Newton shows how to run Art-Net over VPN to overcome some of the protocol's limitations.

\medskip\noindent{\bf Architecture for Control Networks (ACN). } ACN~\cite{JiangJR10} is a set of ANSI standards which define a protocol suite for controlling lighting, networked entertainment devices, and existing control systems. It has been designed to address several shortcomings of existing lighting control protocols, specifically (1) having both an open protocol and specification, (2) media agnostic control, and (3) generalizing to any device that can be controlled \cite{JiangJR10}. ACN defines several protocols on top UDP, and therefore naturally extends to any medium that can carry IP communication. 
There are several implementations of ACN. As an example, OpenACN~\cite{openacn} and the current 
generation of ETCNet~\cite{etcnet} implement the ACN standard.

Security is not addressed in this standard, which assumes that ACN data is transported over a secure network.

\medskip\noindent{\bf Proprietary Protocols. }
Alongside aforementioned lighting control protocols, there are proprietary vendor-centric solutions, 
such as Philips Dynalite and Philips KiNET. 
Philips Dynalite is designed for controlling lighting, interfacing to HVAC, security, fire detection systems, 
and other building sensors \cite{philips-dynet}. The control protocol uses multi-drop serial over 
twisted pair and its characteristics are similar to DMX.
Philips KiNET~\cite{kinet} is a contemporary proprietary lighting protocol based on the 
Philips Color Kinetics platform for LED lighting technology. It uses Ethernet as its communication media. 
To the best of our knowledge, KiNET does not offer any form of authentication or encryption between devices.

\subsection{Common BAS protocols}
In contrast with lighting, BAS protocols interconnect multitudes of sensors and actuators across a building, 
including HVAC, building controls, as well as home and office lighting. The turnkey nature of these 
systems combined with the need to inter-operate in more complex buildings has led to both 
manufacturer-specific protocols and standardization of Internet protocols in order to converge 
and distribute control to these systems. As a result, there are several BAS that have gained 
widespread adoption. We review most prominent solutions: KNX~\cite{KNX-specs} (formerly EIB), 
Fieldbus~\cite{fieldbushistory}, LonTalk/LonWorks~\cite{lontalk-brochure}, and BACNet~\cite{bacnet}. 

\smallskip
\noindent{\bf BACNet. } BACNet~\cite{bacnet} is an open standard specifying a protocol at the backbone 
level to communicate with devices at the control level. Although fixtures
can support BACNet directly, one of the primary goals of BACNet is interoperability 
with other BAS. For the backbone level, various data link and
physical media technologies have been specified based on existing technology, 
such as Ethernet, IP (separately as BACNet/IP), LonTalk, etc.

This standard supports encryption and authentication, although it has been shown 
to be insecure \cite{okabe,nist-bacnet,zachary,GranzerK10,GranzerPK10}. 
DES is the only supported block cipher, and the authentication technique 
is susceptible to man-in-the-middle attacks~\cite{okabe}. Moreover, the protocol 
does not offer protection against interleaving and replay attacks~\cite{okabe}.
In the last decade, there have been several efforts aimed at securing BACNet, 
e.g.~Robin et al.~\cite{robin}.

\medskip\noindent{\bf KNX/EIB. }
KNX began as an open standard -- superseding the legacy European Installation Bus (EIB) -- 
converging several existing standards in home automation and intelligent buildings. Typical KNX 
deployments are used to manage a multitude of building control applications, such as lighting, 
HVAC, energy management and metering.
KNX supports a variety of communication media such as Radio (KNX-RF), Infrared and Powerline.

As outlined in \cite{GranzerKN06,GranzerK10,GranzerPK10}, KNX provides no data security. The control communication to the fixture is limited to a rudimentary access control scheme. The access control scheme uses access levels to define privileges ranging from 0 (highest privilege) to 255 (lowest privilege). Each access level supports a 4-byte password stored and transmitted in clear-text.
Two extensions proposed to improve the security of KNX are EIBsec \cite{GranzerKN06} and a method combining Diffie-Hellman and AES \cite{CavalieriSG09}. 

Cavalieri et al.~\cite{CavalieriSG09} present an extension to the KNX application layer that 
uses a configuration manager placed in the control plane to distribute keys to fixtures 
using Diffie-Hellman~\cite{DiffieH76}. This key exchange is used to establish a secure challenge, followed 
by a challenge-response protocol to authenticate the parties. A long-term key is established for use with 
AES. This proposal specifies no security model, and protocol details are insufficient to evaluate its security. 

\medskip\noindent{\bf LonTalk/LonWorks. }
LonTalk is the communication protocol for the LonWorks BAS. It supports several media types, 
such as RF, Infrared, Coaxial cable, and Fiber Optics. It is a major competitor of KNX and 
is currently in widespread use in building automation.

As specified in \cite{okabe,SchwaigerT03,GranzerKN06,GranzerK10,GranzerPK10}, LonTalk provides minimal 
security. The only security feature is a protocol for data origin authentication in both unicast and 
multicast. Each entity is limited to a single key of up to 48 bits. All entities must share the same key if they 
want to verify messages amongst each other. No security mechanism is provided to distribute keys. 
Thus, each device must be bootstrapped off-line, in a secure environment. Significant overhead is incurred for 
authentication as the protocol requires a 4-round challenge-response protocol
invoked for each message the sender transmits. 

\medskip\noindent{\bf Fieldbus. }
Fieldbus \cite{fieldbushistory} is traditionally used in industrial automation and control. 
It is an open architecture with published IEC standards, which has lead to industry adoption 
and proliferation of several vendors. Its specification defines a communication bus protocol 
for monitoring sensors and operating actuators. The technology has been later adapted to 
building automation and control. 

As outlined by Tretyl et al.~\cite{TretylSS04}, in-band security mechanisms for 
Fieldbus implementations offer weak security. Most implementations offer rudimentary 
access control and authentication where passwords  are sent over  the network in cleartext. 
Security of a typical Fieldbus deployment relies on network isolation. \cite{TretylSS04} also 
proposes the use of standard mechanisms for securing communications in Fieldbus 
deployments, e.g., SSL/TLS and IPSec.

\medskip\noindent{\bf Proprietary Protocols. }
There are several contenders in the BAS ecosystem with proprietary protocols, such as 
Siemens and Honeywell. Since published specifications are scarce we do not overview 
these systems.

\section{Summary and Future Work}
\label{sec:conclusions}
This paper focused on securing instrumented environments connected via Content-Centric Networking (CCN),
motivated by the increasing integration of Building Automation Systems (BAS) with enterprise
networks and the Internet.
In particular, we explore lighting systems over Named-Data Networking (NDN), a 
prominent instance of CCN. 
We identified
security requirements in lighting control and constructed a concrete NDN-based 
security architecture. We then analyzed its security properties and reported on the prototype
implementation and experimental results.  

Clearly, this work represents only the initial step towards assessing suitability of NDN
for communication settings far from its {\em forte} of content distribution. Much more 
work is needed to securely adapt NDN to other types of instrumented environments.
Lighting control is, in some ways, simpler than other BAS types. For example, we assumed
a limited model of feedback in which most of the time command acks are basically one-bit values. This allowed us to 
use tricks based on hash chains or {\em encrypted challenges} to obtain efficiency. In other circumstances,
acks may be more expressive. 

Our current design does not support multicast communication. 
In order for a group of fixtures to be synchronously commanded by the same application, 
each fixture needs to issue a separate interest to the application at roughly the same time. 
The application could then issue a command that would reach all ``interested'' fixtures.

Another direction worth exploring is the utility of long-lived interests, i.e., interests that
establish state in NDN routers but do not expire quickly. This can be useful if we allow
fixtures to issue interests to controlling applications (rather than the other way around
as we do now). A fixture \Fixt\ would issue an interest to \App\ and the latter would only
emit corresponding content when it has a command for \Fixt. This would require \Fixt\ to 
periodically refresh interests as they expire and get flushed by routers.

While not discussed in detail here, NDN offers other significant benefits to BAS applications 
that have motivated this exploration, such as providing network access to sensing and 
actuation points  via application-assigned data names, without the need to specify
IP host addresses and port numbers for gateways. 

\newpage

\bibliographystyle{abbrv}
\bibliography{references}

\begin{thebibliography}{10}

\bibitem{bacnet}
ANSI.
\newblock Standard 135-1995, {BACnet} a data communication protocol for
  building automation and control networks, 1995.

\bibitem{artnet}
{Specification for the {Art-Net 3} {E}thernet Communication Standard}.
\newblock \texttt{http://www.artisticlicence.com}.
\newblock Retrieved Feb. 2012.

\bibitem{asokan}
N.~Asokan, G.~Tsudik, and M.~Waidner.
\newblock Server-supported signatures.
\newblock {\em Journal of Computer Security}, 5:131--143, 1996.

\bibitem{CavalieriSG09}
S.~Cavalieri and G.~Cutuli.
\newblock Implementing encryption and authentication in {KNX} using
  {Diffie-Hellman} and {AES} algorithms.
\newblock In {\em IEEE IECON}, pages 2459--2464, Nov. 2009.

\bibitem{ChallalBH04}
Y.~Challal, A.~Bouabdallah, and Y.~Hinard.
\newblock Efficient multicast source authentication using layered hash-chaining
  scheme.
\newblock In {\em IEEE LCN}, Nov. 2004.

\bibitem{andana}
S.~DiBenedetto, P.~Gasti, G.~Tsudik, and E.~Uzun.
\newblock {ANDaNA}: Anonymous named data networking application.
\newblock In {\em NDSS}, 2012.

\bibitem{DiffieH76}
W.~Diffie and M.~Hellman.
\newblock New directions in cryptography, 1976.

\bibitem{DingMT02}
X.~Ding, D.~Mazzocchi, and G.~Tsudik.
\newblock Experimenting with server-aided signatures.
\newblock In {\em NDSS}, 2002.

\bibitem{dmx}
{DMX512-A}.
\newblock \texttt{http://www.opendmx.net/index.php/DMX512-A}.
\newblock Retrieved Feb. 2012.

\bibitem{dmx-over-rf}
Lighting systems made easy a guide to lighting installations.
\newblock
  \texttt{http://www.leprecon.com/catalogs/\\280075BLightingMadeEasy.pdf}.
\newblock Retrieved Feb. 2012.

\bibitem{philips-dynet}
An introduction to the {Philips Dynalite} control system.
\newblock
  \texttt{http://www.lighting.philips.com/pwc\_li/main/\\subsites/dynalite/library\_support/assets/\\general\_brochures/intro\_to\_phd\_controls\_systems.pdf}.
\newblock Retrieved Feb. 2012.

\bibitem{etcnet}
{Electronic Theater Controls (ETCNet)}.
\newblock \texttt{http://www.etcconnect.com/}.
\newblock Retrieved Feb. 2012.

\bibitem{fieldbushistory}
{Fieldbus} history.
\newblock \texttt{http://www.fieldbus.org/}.
\newblock Retrieved Feb. 2012.

\bibitem{DBLP:conf/asiacrypt/GentryS02}
C.~Gentry and A.~Silverberg.
\newblock Hierarchical id-based cryptography.
\newblock In {\em ASIACRYPT}, pages 548--566, 2002.

\bibitem{GranzerK10}
W.~Granzer and W.~Kastner.
\newblock Security analysis of open building automation systems.
\newblock In {\em SAFECOMP}, pages 303--316, 2010.

\bibitem{GranzerKN06}
W.~Granzer, W.~Kastner, G.~Neugschwandtner, and F.~Praus.
\newblock Security in networked building automation systems.
\newblock In {\em IEEE WFCS}, pages 283--292, Jun. 2006.

\bibitem{GranzerPK10}
W.~Granzer, F.~Praus, and W.~Kastner.
\newblock Security in building automation systems.
\newblock {\em IEEE IES}, 57(11):3622--3630, Nov. 2010.

\bibitem{gritter2001architecture}
M.~Gritter and D.~Cheriton.
\newblock An architecture for content routing support in the internet.
\newblock In {\em USENIX USITS}, 2001.

\bibitem{gumstix}
{Gumstix Overo Air}.
\newblock
  \texttt{http://www.gumstix.com/store/\\product\_info.php?products\_id=226}.
\newblock Retrieved Feb. 2012.

\bibitem{nist-bacnet}
D.~Holmberg and D.~Evans.
\newblock Bacnet wide area network security threat assessment.
\newblock \texttt{http://fire.nist.gov/bfrlpubs/build03/art034.html}, Jul.
  2003.

\bibitem{Hu05efficientconstructions}
Y.~Hu, M.~Jakobsson, and A.~Perrig.
\newblock Efficient constructions for one-way hash chains.
\newblock In {\em ACNS}, pages 423--441, 2005.

\bibitem{Huntington}
J.~Huntington.
\newblock {\em Control Systems for Live Entertainment, 3rd Edition}.
\newblock Burlington, MA: Focal Press, 2007.

\bibitem{VanSmBriPlStThBra09-Voice}
V.~Jacobson, D.~Smetters, N.~Briggs, M.~Plass, J.~Thornton, and R.~Braynard.
\newblock {VoCCN}: Voice-over content centric networks.
\newblock In {\em ReArch}, 2009.

\bibitem{Jacobson2009}
V.~Jacobson, D.~Smetters, J.~Thornton, M.~Plass, N.~Briggs, and R.~Braynard.
\newblock Networking named content.
\newblock In {\em ACM CoNEXT}, 2009.

\bibitem{JiangJR10}
W.~Jiang, Y.~Jiang, and H.~Ren.
\newblock Analysis and prospect of control system for stage lighting.
\newblock In {\em IEEE CISP}, volume~8, pages 3923--3929, Oct. 2010.

\bibitem{kinet}
Philips multi-protocol converter.
\newblock
  \texttt{http://www.colorkinetics.com/\\support/datasheets/\\Multi-Protocol\_Converter\_SpecSheet.pdf}.
\newblock Retrieved Feb. 2012.

\bibitem{KNX-specs}
{KNX} system specifications v 3.0.
\newblock \texttt{http://www.knx.org//downloads-support/downloads/}.

\bibitem{koponen2007data}
T.~Koponen, M.~Chawla, B.~Chun, A.~Ermolinskiy, K.~Kim, S.~Shenker, and
  I.~Stoica.
\newblock A data-oriented (and beyond) network architecture.
\newblock In {\em ACM SIGCOMM}, volume~37, pages 181--192. ACM, 2007.

\bibitem{lontalk-brochure}
{i.LON} 600 {LonWorks/IP} server models 760x.
\newblock
  \texttt{http://www.echelon.com/support/documentation/\\datashts/7260x.pdf}.
\newblock Retrieved Feb. 2012.

\bibitem{NDN}
Named data networking project {(NDN)}.
\newblock \texttt{http://named-data.org}.
\newblock Retrieved Feb. 2012.

\bibitem{newton05}
S.~Newton.
\newblock Art-net and wireless routers.
\newblock In {\em IEEE APCC}, pages 857 --861, Oct. 2005.

\bibitem{okabe}
N.~Okabe, S.~Sakane, K.~Miyazawa, K.~Kamada, A.~Inoue, and M.~Ishiyama.
\newblock Security architecture for control networks using {IPsec} and {KINK}.
\newblock In {\em IEEE SAINT}, pages 414--420, 2005.

\bibitem{openacn}
{Open ACN} project.
\newblock \texttt{http://openacn.engarts.com}.
\newblock Retrieved Feb. 2012.

\bibitem{robin}
D.~Robin.
\newblock Internet security protocols for {BACnet}, {BACnet} committee {SSPC}
  135 document number {DR-028-1}, 2002.

\bibitem{SchwaigerT03}
C.~Schwaiger and A.~Treytl.
\newblock Smart card based security for fieldbus systems.
\newblock In {\em IEEE ETFA}, volume~1, pages 398--406, Sept. 2003.

\bibitem{ndn-encryption}
D.~Smetters.
\newblock {CCNx} access control specifications.
\newblock Technical report, PARC, 2010.

\bibitem{TretylSS04}
A.~Treytl, T.~Sauter, and C.~Schwaiger.
\newblock Security measures for industrial fieldbus systems - state of the art
  and solutions for {IP}-based approaches.
\newblock In {\em IEEE WFCS}, pages 201--209, Sept. 2004.

\bibitem{YavuzN09}
A.~Yavuz and P.~Ning.
\newblock Hash-based sequential aggregate and forward secure signature for
  unattended wireless sensor networks.
\newblock In {\em ICST Mobiquitous}, pages 1--10, Jul. 2009.

\bibitem{zachary}
J.~Zachary, R.~Brooks, and D.~Thompson.
\newblock Secure integration of building network into the global internet.
\newblock \texttt{http://fire.nist.gov/bfrlpubs//build03/art027.html}, Oct.
  2002.

\end{thebibliography}
\end{document}